\documentclass[aps,prl,reprint,amsfont,amsmath,amssymb,floatfix]{revtex4-2}
\usepackage{graphicx,bm,xcolor}
\usepackage[normalem]{ulem}
\usepackage{hyperref}
\usepackage{orcidlink}
\hypersetup{colorlinks=true,allcolors=blue}

\begin{document}

\title{Kinetic inductance of two-dimensional viscous electron fluid} 

\author{A.~A.~Grigorev\,\orcidlink{0009-0009-6614-7287}}
\affiliation{Ioffe Institute, St.~Petersburg 194021, Russia}
\author{ A.~N.~Afanasiev\,\orcidlink{0000-0002-8782-2044}}
\email{afanasiev.an@mail.ru}
\affiliation{Ioffe Institute, St.~Petersburg 194021, Russia}


\begin{abstract}
Modern high-mobility two-dimensional electron systems are pure enough to manifest hydrodynamic regime of electron transport similar to dynamics of classical fluids. With the bulk scattering controlled by momentum-conserving electron-electron collisions, these novel hydrodynamic materials demonstrate systematic violation of conventional metallic behavior due to the Drude model in stationary electric transport. In this letter, we study the linear reactive AC response of 2D electron fluid in a narrow channel geometry and show that the essentially non-local viscous dissipation induces non-trivial size effect in it. In particular, kinetic inductance, which represents collective inertia of the entire electron fluid, is enhanced by non-uniform flow profile and, in contrast to diffusive transport, dominates the electrical impedance already in collisional regime by virtue of viscous skin effect. The onset of reactive behavior falls within the microwave frequency domain in the state-of-the-art 2D hydrodynamic conductors which makes them promising material system for applications in high-frequency electronics at elevated temperatures.
\end{abstract}

\maketitle


{\bf 1. Introduction.}
Drude model provides simple yet powerful framework to describe long-wavelength quasi-classical transport properties of conducting materials. Developed in the early years of condensed matter physics, this theory has become paradigmatic and its predictions are relevant for the majority of conventional metallic systems. The latter is due to the fact that in most conductors electron transport regime is {\it diffusive} by virtue of intense momentum-relaxing scattering by phonons and impurities in the bulk. Rare electron-electron collisions give minor contribution to transport properties in diffusive regime and are typically ignored. Furthermore, local DC conductivity of a single component Galilean-invariant electron system remains unaffected by interparticle scattering~\cite{Landau1936, Pal2012, Kovalev2025} (excluding umklapp processes) due to momentum and hence current conservation during collisions.

The opposite scenario of frequent electron-electron collisions which dominate over momentum-relaxing scattering, namely $\ell_{\rm ee}\ll\ell_{\rm p}$ (here $\ell_{\rm ee,p}$ are corresponding mean free paths), was considered exotic and hardly feasible for a long time. Nevertheless, the ongoing search for a proper material to meet this condition was motivated by unusual transport properties of such metallic systems, first considered in the pioneering paper by R.~N.~Gurzhi~\cite{Gurzhi1963}. Intense interparticle scattering rapidly relaxes any perturbation to the local equilibrium state described by macroscopic drift velocity ${\bf V}({\bf r},t)$, chemical potential $\mu({\bf r}, t)$ and electron temperature $T({\bf r},t)$. As a result, dynamics of this collectively behaving electron system resembles the one of classical fluids~\cite{Narozhny2019, Varnavides2023, Hui2025}. Instead of the Drude model, long-wavelength charge transport in this {\it hydrodynamic} regime is described by the Navier-Stokes-Ohm equation~\cite{Hui2025}.

Practical realization of electron hydrodynamics has come with advances~\cite{Dean2010, Wang_1D_contact_2013, Mackenzie2017, Chung2021} in technology of ultraclean two-dimensional electron systems (2DES) about a decade ago~\cite{Bandurin2016, Crossno2016, Moll2016}. Since then, fluid-like electron transport has been demonstrated in a number of high-quality 2D materials, graphene~\cite{Bandurin2016, Crossno2016, Sulpizio2019, Ku2020, Huang2023, Palm2024, Talanov2024} and GaAs quantum wells~\cite{Gusev2018, Gusev2021, Wang2022, Vijayakrishnan2023, Sarypov2025} being the most popular and well-studied among them. The necessary condition $\ell_{\rm ee}(T,N_0)\ll \ell_{\rm p}(T)$ is satisfied in extrinsic graphene at low-temperature electron mobilities of the order of $10^5-10^6$ cm$^2$/V$\cdot$s~\cite{Sulpizio2019, Huang2023} and at $10^6-10^7$ cm$^2$/V$\cdot$s in GaAs nanostructures~\cite{Gusev2018, Braem2018, Wang2022}. The hydrodynamic window is opened up at lattice temperatures $T=100-250$~K~\cite{Ho2018} and $T=10-40$~K~\cite{Ahn2022}, respectively, and moderate carrier densities $N_0$. In the finite size samples there is an additional requirement for hydrodynamic transport that electron-electron scattering should also surpass momentum relaxation at the boundaries, $\ell_{\rm ee}(T,N_0)\ll W$, with $W$ the characteristic size of constriction, typically of the order of several microns.

Transport in two-dimensional electron fluids (2DEF) significantly deviates from conventional metals and moderate mobility 2DES. One of the hallmarks of electron hydrodynamics is violation of the Wiedemann-Franz law~\cite{Principi2015, Crossno2016, Ahn2022, Talanov2024}. This is a direct consequence of electron-electron scattering affecting charge and thermal current relaxation in different ways. Another group of unconventional transport phenomena is mesoscopic finite size effects~\cite{Sondheimer1952, Molenkamp1994}. In the case of 2DEF, these are manifestations of essentially non-local viscous dissipation which controls transport in the bulk alongside with momentum relaxation at the boundaries. In particular, electron viscosity is responsible for the non-uniform flow profiles of 2DEF~\cite{Sulpizio2019, Ku2020}, formation of current whirlpools and vortices~\cite{AharonSteinberg2022, Palm2024,Shevyrin2026}, size-dependent resistivity~\cite{Moll2016, Alekseev2016, Wang2022, Huang2023}, Johnson-Nyquist noise~\cite{Hui2023} in hydrodynamic samples and the Gurzhi effect~\cite{Gurzhi1963, Gusev2021, Vijayakrishnan2023}. In perpendicular magnetic field, the viscosity effect is also non-trivial as it leads to the giant negative magnetoresistance~\cite{Alekseev2016, Gusev2018, Sulpizio2019, Berdyugin2019, Wang2022} and anomalous contribution to the Hall effect~\cite{Scaffidi2017, Gusev2018, Berdyugin2019}. At the present time, 2D hydrodynamic conductors are primarily considered~\cite{Varnavides2023, Huang2024} as promising materials for energy-efficient nanoelectronic devices, which offer significant reduction of channel resistance.

For non-stationary transport in high-mobility 2DES, collision-dominated (hydrodynamic) behavior is relevant at frequencies below the electron-electron scattering rate $\omega\leq \nu_{\rm ee}$, which marks the onset of collisionless (ballistic) regime. In ultrapure samples of GaAs and graphene, the characteristic values of $\nu_{\rm ee}$, 0.1 THz and 1 THz, respectively, belong to far infrared domain. Studies of high-frequency properties of 2DEF in graphene have been primary focused on tracing bulk collective modes across this hydrodynamic-to-ballistic crossover~\cite{Lucas2018, Svintsov2018, Kapralov2022, BarconsRuiz2023}. In the case of GaAs structures, the main object of interest is the viscoelastic behavior of 2DEF, induced by short-range Landau interaction, and the associated transverse shear sound waves~\cite{Alekseev2019, Wang2022, Afanasiev2023}. 

\begin{figure}[t!]
\centerline{\includegraphics[width=.99 \linewidth]{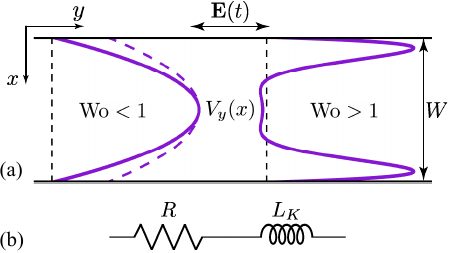}}
\caption{
Schematic representation of the AC electron flow in a narrow channel.
(a) Sketch of the system's geometry and the longitudinal flow velocity profile $V_y(x)$. Solid purple curves correspond to a no-slip flow and show the real part of $V_y(x)$ in the quasi-static Poiseuille regime (${\rm Wo}< 1$) and the high-frequency Womersley regime (${\rm Wo}> 1$) when the viscous skin layers are developed. Dashed purple curve shows the effect of a finite slip length on the Poiseuille flow and dashed black lines mark $V_y(x)=0$.  (b) AC equivalent electric circuit for ungated 2D electron system.
\label{Fig1}
}
\end{figure}

By contrast, incoherent AC transport in 2DEF in finite size constrictions is studied in less detail. Direct calculation of the AC flow admittance in the most popular narrow channel geometry~\cite{Varnavides2023} showed no qualitative differences in its frequency scaling in comparison to the ohmic transport~\cite{Moessner2018}. Nevertheless, evolution of the hydrodynamic velocity profile with increasing $\omega$ is non-trivial~\cite{Womersley1955, Afanasiev2022}, see Fig.~\ref{Fig1}a. At low frequencies, the flow resembles quasi-static Poiseuille shape. At higher $\omega$, the balance between non-local viscous and local inertial forces shifts towards the latter. Consequently, the flow profile undergoes pronounced transformation: in the center of a channel it is almost flat and reactive, while the dissipative contribution to ${\bf V}({\bf r},t)$, which is in-phase with the driving uniform electric field ${\bf E}(t)$, is pushed to the boundaries. The developed viscous skin layers~\cite{Landau_V6} are characteristic feature of the Womersley regime~\cite{Womersley1955}, first considered for the alternating blood flow in large-radius arteries of vertebrate animals.

From the perspective of high-frequency electronics, linear AC response of ungated 2DES can be represented by the equivalent circuit shown in Fig.~\ref{Fig1}b. Here $R$ is the sample resistance responsible for the active component of a net current due to dissipation in the system, while the reactive response is captured by the kinetic inductance $L_{\rm k}$~\cite{Burke2000, Kang2005, Andress2012, Yoon2014, Chee2025} related to collective inertia of charge carriers. Within the Drude model, local complex resistivity of non-interacting electrons takes the form $\rho_{_{\rm D}}(\omega)=\rho_{\rm_0}-i\omega \lambda_{\rm k}$, hence the circuit components are given by the static Drude resistivity $\rho_0=m/e^2N_0\tau_{\rm p}$ and the specific kinetic inductance $\lambda_{\rm k}=m/e^2 N_0$. Here $e$ is the electron charge and $m$ is the mass of inertial density of 2DES (also dubbed as collective or plasmonic mass~\cite{Yoon2014}). Correspondingly, the onset of reactive behavior identified by the kinetic inductor's quality factor $Q_{\rm k}=\omega L_{\rm k}/R$ equal to unity coincides with collision frequency $\omega=\nu_{\rm p}$. In high-mobility 2DES, the temperature dependence of the latter due to electron-phonon scattering results in predominantly resistive response and collisionally overdamped collective modes up to terahertz frequencies at room temperature~\cite{Dzhikirba2023}. Therefore, reactive behavior of these structures in microwave (MW) domain and their compatibility with electronics is traditionally considered only at cryogenic temperatures~\cite{Burke2000, Kang2005, Andress2012, Yoon2014, Chee2025}.
At intermediate temperatures within the electron hydrodynamics window, the transport in 2DES on one side is also characterized by the far infrared collision frequency $\nu_{\rm ee}$, on the other side inductive phase delays between current and voltage~\cite{Womersley1955, Moessner2018} are present locally at $\omega<\nu_{\rm ee}$ in constrained geometries. The important issue of reactive properties of 2DEF in finite size flows remains unresolved up to now and is, therefore, the primary goal of the present work.

In this Letter, we revisit the linear AC transport of 2D hydrodynamic electrons with the emphasis on its reactive component. First, we consider an alternating incoherent flow of ungated 2DEF in narrow channel geometry and study the properties of its equivalent kinetic inductor. We show that in the presence of non-local viscous dissipation the reactive behavior of 2DES is enhanced: its onset shifts deep into collisional regime towards MW frequencies and the kinetic inductance is increased with respect to $\lambda_{\rm k}$. Next, for gated narrow channel geometry we show that reactive properties of 2DEF are imprinted in propagation of acoustic plasmons below $\nu_{\rm ee}$ and their stability against collisional damping. This behavior can be directly probed by MW transmission in purely electronic setup~\cite{Andress2012, Yoon2014, Chee2025}. Finally,  we recall an energy argument and apply thermodynamics of the local equilibrium state to derive expression for $L_{\rm k}$ of arbitrary parallel flow of 2DEF which gives a guideline for manipulation of $L_{\rm k}$ via geometric engineering of 2DEF flows in device applications.

\vspace{0.1cm}
{\bf 2. Reactive response of 2D viscous electron fluid in a narrow channel.} To study reactive response  of 2DEF, first, we focus on the narrow channel geometry shown in Fig.\ref{Fig1}a and consider an oscillating linear flow driven by longitudinal uniform electric field ${\bf E}(t)={\bf E}_0e^{-i \omega t}$ in ungated system. This setup corresponds to the case when the channel length is small compared to bulk plasmon wavelength at driving frequency $\omega$ and plasmon excitation is suppressed. Hence the low amplitude dynamics of 2DEF is incoherent and incompressible. Therefore, fluid motion in the channel is described by longitudinal flow velocity $V_y(x,t)$ which is non-uniform only in transverse direction, see Fig.~\ref{Fig1}a.  Furthermore, we consider the low-temperature limit $T\ll E_{\rm F}$ which is typical for hydrodynamic transport in 2DES.

Under these conditions, incompressible viscous parallel flow of 2DEF is described by the linearized non-stationary Navier-Stokes equation for the complex amplitude of flow velocity $V_y(x,t)=V(x) e^{-i\omega t}$:
\begin{equation}
    \label{Eq:lin_NS}
    -i\omega V=\eta\partial_x^2 V -e E_0/m\,,
\end{equation}
where $\eta=v_{\rm F}\ell_{\rm ee}/4$ is the kinematic shear viscosity and $\ell_{\rm ee}$ is the shear stress relaxation mean free path due to electron-electron collisions. We note that for collision-dominated AC flows ($\omega \tau_{\rm ee}\ll 1$), the viscoelastic behavior of electron fluid~\cite{Conti1999} described by frequency-dependent and complex-valued shear viscosity $\eta(\omega)$, can be ignored.

In 2DEF, dissipation in the system is controlled by viscosity rather than bulk momentum relaxation, hence the Drude transport equation is replaced by~(\ref{Eq:lin_NS}). For a purely viscous transport, carrier momentum is relaxed solely at the channel boundaries via elastic uncorrelated scattering at current impenetrable walls. In general case, this leads to a finite slip of fluid flow at boundaries~\cite{Kiselev2019, Raichev2022}, described by the boundary conditions
\begin{equation}
    \label{Eq:slip_bc}
    (1-\xi {\bf n}\bm{\nabla})V(\pm W/2)=0\,.
\end{equation}
Here $W$ is the channel width, $\bf n$ stands for the inward normal to the boundary ($n_x=\mp 1$ at $x=\pm W/2$), $\xi=(3\pi \eta/4 v_{\rm F}) (1+r)/(1-r)$ is the slip length and $0\leq r<1$ is the boundary specular reflection coefficient. In this work we use the most common Maxwell approximation for $\xi$~\cite{Raichev2022}. 

\begin{figure}[t!]
\centerline{\includegraphics[width=.99 \linewidth]{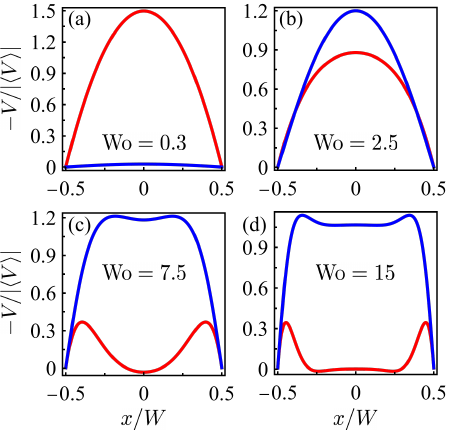}}
\caption{Evolution of the non-uniform drift velocity profile in an AC no-slip channel flow of 2DEF with frequency, measured by the Womersley number ${\rm Wo}(\omega)$. Red and blue curves denote real and imaginary parts of $V(x)$, respectively, scaled to an absolute value of mean velocity over the channel.
\label{Fig2}
}
\end{figure}

Solution of the transport equation~(\ref{Eq:lin_NS}) with boundary conditions~(\ref{Eq:slip_bc}) yields
\begin{equation}
    \label{Eq:V}
    V(x)=-i\frac{e E_0}{m \omega}\left[1-\frac{\cosh(\varkappa x)/\cosh(\varkappa W/2)}{1+\varkappa \xi\tanh(\varkappa W/2)}\right]\,,
\end{equation}
where $\varkappa=\sqrt{-i\omega/\eta}=(1-i)\ell_s^{-1}(\omega)$ is the spatial eigenvalue of~(\ref{Eq:lin_NS}) responsible for viscous skin effect and $\ell_{\rm s}(\omega)=\sqrt{2\eta/\omega}$ is the associated skin depth. The latter is proportional to velocity diffusion length over the period of ${\bf E}(t)$. Evolution of the AC flow profile~(\ref{Eq:V}) with frequency is controlled by the Womersley number ${\rm Wo}(\omega)=W/\ell_{\rm s}(\omega)\sim\sqrt{\omega}$, see Fig.~\ref{Fig2}a-d. Here $\langle\ldots\rangle$ denotes averaging over the channel width. Namely, ${\rm Wo}\sim 1$ (Fig.~\ref{Fig2}b) marks the crossover between the quasi-static resistive Poiseuille regime (Fig.~\ref{Fig2}a) at ${\rm Wo}\ll1$ and reactive Womersley flow (Fig.~\ref{Fig2}d) at ${\rm Wo}\gg 1$. It is instructive to note that viscous skin effect influences only current density distribution, but not the driving field ${\bf E}(t)$. As a result, the AC flow velocity spans over the whole channel at any frequency, while it is the resistive part of $V(x)$ which is expulsed from the bulk. As it is seen from Fig.~\ref{Fig2}, the increase of ${\rm Wo}$ indicates redistribution of the local phase of current density. We also note that the upper bound for the Womersley number imposed by the hydrodynamic-to-ballistic crossover frequency $\omega=\nu_{\rm ee}$ is ${\rm Wo}(\nu_{\rm ee})=\sqrt{2}/{\rm Kn}$, where ${\rm Kn}=\ell_{\rm ee}/W$ is the Knudsen number, which indicates how deep 2DES is in the hydrodynamic regime. Namely, the ${\rm Kn}\to 0$ limit corresponds to ideal fluid, at finite ${\rm Kn}<1$ 2DEF is viscous and ${\rm Kn}=1$ marks the hydrodynamic-ballistic crossover in stationary transport.

\begin{figure*}[t!]
\centerline{\includegraphics[width=.99 \linewidth]{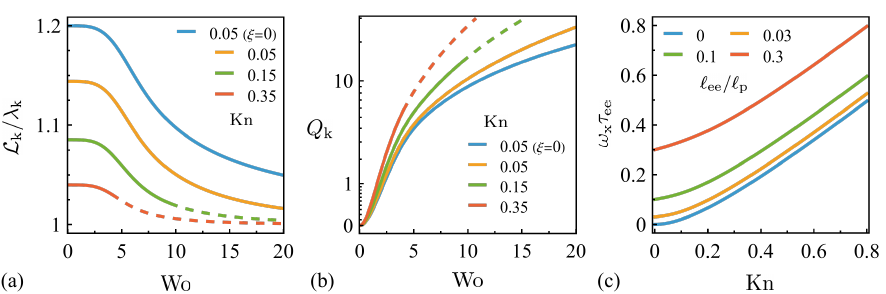}}
\caption{
Linear reactive response of viscous 2DEF for a narrow channel flow. (a) Enhancement of kinetic inductance at various frequencies, measured by the Womersley number ${\rm Wo}(\omega)$, and flow slip, measured by the Knudsen number ${\rm Kn}$. (b) Kinetic inductor's quality factor at different ${\rm Wo}$ and ${\rm Kn}$. (c) Resistive-to-reactive crossover frequencies for different states of viscous 2DEF with weak momentum relaxation. Dashed parts in (a) and (b) correspond to formal extrapolations to the collisionless domain $\omega\tau_{\rm ee}>1$; results for the no-slip case are denoted by $\xi=0$.
\label{Fig3}
}
\end{figure*}

Expression for two-terminal sheet impedance of the channel $\mathcal{Z}(\omega)=E_0 W/I$ follows readily from Eq.~(\ref{Eq:V}) and the net current complex amplitude $I=-e N_0 W\langle V\rangle$, which give
\begin{equation}
    \label{Eq:Z}
    \mathcal{Z}(\omega)=-i\omega\lambda_{\rm k}\left[1-\frac{2}{\varkappa W}\frac{\tanh(\varkappa W/2)}{1+\varkappa \xi \tanh(\varkappa W/2)}\right]^{-1}\,.
\end{equation}
The corresponding sheet kinetic inductance of the channel flow of 2DEF $\mathcal{L}_{\rm k}(\omega)=-{\rm Im}\,\mathcal{Z}(\omega)/\omega$ is presented in Fig.~\ref{Fig3}a. It is clearly seen that conventional metallic behavior is also violated in linear reactive response of 2DEF: the kinetic inductance of collective flow is enhanced with respect to $\lambda_{\rm k}$ and demonstrate pronounced size effect via dependence on the Womersley and the Knudsen numbers. The latter measures significance of the flow slip, $\xi/W\sim {\rm Kn}$. Here we consider the case of purely diffusive boundary reflection ($r=0$), however the effect of finite boundary specularity ($r\neq 0$) is equivalent to increasing ${\rm Kn}$ in Figs.~\ref{Fig3}a and~\ref{Fig3}b. Fig.~\ref{Fig3}a demonstrates another important feature: kinetic inductance of 2DEF reduces to $\lambda_{\rm k}$ with increasing frequency and flow slip. Moreover, this occurs simultaneously with flattening of the flow velocity profile towards the uniform shape similar to the Drude case, see Figs.~\ref{Fig1} and~\ref{Fig2}. The effect of slip length on $\mathcal{L}_{\rm k}$ is pronounced. Kinetic inductance enhancement is suppressed already at ${\rm Kn}\approx0.5$. In contrast, a considerable reduction of $\mathcal{L}_{\rm k}$ with increasing frequency starts at relatively high values of the Womersley number (${\rm Wo}>15$), which approach or even lie beyond the upper bound ${\rm Wo}(\nu_{\rm ee})$.

Next, using~(\ref{Eq:Z}) we calculate the quality factor of kinetic inductor $Q_{\rm k}(\omega)=\omega\mathcal{L}_{\rm k}(\omega)/\mathcal{R}(\omega)$, where $\mathcal{R}(\omega)={\rm Re}\,\mathcal{Z}(\omega)$ is the sheet resistance. Fig.~\ref{Fig3}b, shows that in 2DEF the crossover to reactive behavior occurs due to viscous skin effect approximately at ${\rm Wo}\sim 1$, i.e. well below the collision frequency $\nu_{\rm ee}$ at ${\rm Kn}\ll1$. 
This practically important feature of AC transport in 2DEF is in contrast to conventional 2DES. The crossover frequency for the purely viscous fluid estimated as $\nu_{\rm x}\approx \nu_{\rm ee}{\rm Kn}^2/4\pi$ at ${\rm Kn}\ll 1$ is parametrically small compared to $\nu_{\rm ee}$. However, in real hydrodynamic materials the lower bound for $\nu_{\rm x}$ at ${\rm Kn}\ll 1$ is set by the rate of weak and secondary bulk momentum relaxation $\nu_{\rm p}$. The latter is taken into account in Eqs.~(\ref{Eq:lin_NS}), (\ref{Eq:V}) and (\ref{Eq:Z}) by introducing shift $\omega\to\omega+i\nu_{\rm p}$. This produces another characteristic scale - the Gurzhi length $\ell_{\rm G}$ - which describes velocity diffusion over the momentum relaxation time $\tau_{\rm p}$ and the associated dimensionless Gurzhi number ${\rm Gu}=W/\ell_{\rm G}$. The flow of 2DEF is purely viscous at ${\rm Gu}< 1$ while at ${\rm Gu}>1$ the flow is non-uniform only in the near-boundary viscous layers of width $\ell_{\rm G}$. Fig.\ref{Fig3}c shows the crossover frequencies in 2DEF according to equation $Q_{\rm k}({\rm Wo}_{\rm x}, {\rm Kn}, {\rm Gu})=1$ numerically solved for ${\rm Wo}_{\rm x}={\rm Wo}(\omega_{\rm x})$. Growing nonparabolicity in $\omega_{\rm x}({\rm Kn})$ with increasing ${\rm Kn}$ is the effect of flow slip. This feature is enhanced with boundary specularity $r$ with the overall curves in Fig.~\ref{Fig3}c becoming flatter. In Fig.~\ref{Fig3}, ${\rm Kn}=0.05$ and $\ell_{\rm ee}/\ell_{\rm p}=0.03$ correspond to the smallest reported parameters of 2DEF flow~\cite{Sulpizio2019}. For this case, crossover occurs at $\omega_{\rm x}\tau_{\rm ee}=0.04$, leading to MW threshold frequencies of 1.3~GHZ in GaAs ($\nu_{\rm ee}=0.2\,{\rm THz}$) and 13~GHz in graphene ($\nu_{\rm ee}=2\,{\rm THz}$). Therefore, resistive-to-reactive crossover in 2DEF can be directly probed at elevated temperatures by the phase of MW reflection coefficient from mesoscopic 2DES in purely electrical setup~\cite{Burke2000, Kang2005}.


{\bf 3. Propagation of acoustic plasmons in gated 2D viscous electron fluid in a narrow channel.} Taking into account the enhanced reactivity of ungated 2DEF in incoherent AC flow at MW frequencies, we now turn to study how this behavior manifests in plasmon propagation along the gated narrow channel in geometry similar to Fig.~\ref{Fig1}. This setup corresponds to direct launching of plasmons over a continuous MW frequency range~\cite{Andress2012, Yoon2014, Chee2025} in structures where narrow channel of high-mobility 2DES is merged with the signal lines of an input and output coplanar waveguides and gate serves as MW ground. Plasmon dispersion is then measured by the propagation phase delay $\phi(\omega)={\rm Re}\,k(\omega)$ in transmission of MW signal at frequency $\omega$. Hereinafter $k(\omega)$ is the longitudinal wave vector.

The typical design of these structures utilizes proximate gate at distance $d\ll W$ to slow down screened acoustic plasmon (AP) velocity to $s=10^8-10^9$ cm/s. With the characteristic channel widths of several micrometers in both electron hydrodynamics structures and plasmon waveguides in Refs.~\cite{Andress2012, Yoon2014, Chee2025}, the bulk AP wavelength at MW is the longest among spatial scales in the system $k^{-1}_0(\omega)\gg W \gg d,\ell_{\rm ee}$, except for the channel length ($k_0=\omega/s$). Provided that actual plasmon wave vector $k(\omega)$ satisfies the same conditions, only gapless quasi-1D mode is excited in this narrow strip 2D plasmon waveguide~\cite{Rudin1997, Satou2005} and its propagation is determined by collisions and strong screening by the gate. The latter justifies application of the local capacitance approximation (LCA) $\varphi(x)=-e\delta N(x)/C$ for the self-consistent potential $\varphi({\bf r},t)=\varphi(x)\,e^{i(ky-\omega t)}$, which ignores edge electrostatic effects near the boundaries and describes the plasmon dispersion to the leading order by $kd\ll 1$~\cite{Nazin1987,Volkov1988}. Here $\delta N(x)$ is amplitude of oscillating electron density in 2DEF and $C=\epsilon/4\pi d$ is the channel-gate capacitance with $\epsilon$ the background dielectric constant.

Coherent oscillations in compressible flow of 2DEF in narrow channel are described by the system of linearized continuity and Navier-Stokes equations with LCA in-plane self-consistent field ${\bf E}({\bf r},t)=-{\bm \nabla}\varphi({\bf r},t)$, accompanied by the slip~(\ref{Eq:slip_bc}) and open circuit $V_{x}(\pm W/2)=0$ boundary conditions for the drift velocity components along and across the channel, respectively. Since the typical AP velocities in experiments satisfy $s\gg v_{\rm F}$, we neglect the first sound contribution to $s$ which is then given by $s=1/\sqrt{\lambda_{\rm k}C}$ in terms of transmission line model for local and lossless transport in 2DES~\cite{Burke2000, Kang2005, Andress2012, Yoon2014, Chee2025}.

Bulk eigenmode spectrum of the governing system of equations given by
\begin{gather}
    \label{Eq:Bulk_AP}
    k_{\rm pl}^2=\omega(\omega+i\nu_{\rm p})/(s^2-i\omega\eta)\,,\\
    \label{Eq:Bulk_Shear}
    k_{\eta}^2=(i\omega-\nu_{\rm p})/\eta\,,
\end{gather}
consists of longitudinal AP mode~(\ref{Eq:Bulk_AP}) and transverse diffusive shear mode~(\ref{Eq:Bulk_Shear}), which are coupled in the finite size case. The long wavelength condition $k\ll k_{\eta}, W^{-1}$ together with $s^2 \gg \omega \eta$ leads to $k,q_{\rm pl}\ll q_{\eta}, W^{-1}$, where $q_{{\rm pl},\eta}^2=k_{{\rm pl},\eta}^2-k^2$ are transverse components of 2D wave vectors ${\bf k}_{{\rm pl},\eta}$. Therefore, at small $k$ dissipative transverse shear modes are excited at the boundaries across the channel towards its center and are coupled only to the longitudinal field $E_{y}$ of plasmon through $V_y$. Since the self-consistent field varies slowly across the channel width and the shear mode wavelength, viscous response to plasmon propagation is described by Eqs.~(\ref{Eq:V}) and~(\ref{Eq:Z}) giving for the complex amplitude of the longitudinal net current $I_y=E_y W/\mathcal{Z}(\omega+i\nu_{\rm p})$. Combining it with the self-consistent expression for the plasmon field due to LCA and continuity equation $E_y=k^2 I_y/(i\omega C W)$, we obtain the dispersion equation
\begin{equation}
    \label{Eq:Disp}
    k(\omega)=\omega\sqrt{\mathcal{L}_{\rm k}(\omega)C\,[1+i/Q_{\rm k}(\omega)]}\,,
\end{equation}
which generalizes lossy transmission line model for plasmons in 2DES for the case of non-local transport in narrow channel. According to~(\ref{Eq:Disp}), the long wavelength limit mentioned above is guarantied by general hydrodynamic regime requirements and additional condition $s/v_{\rm F}>\sqrt{\ell_{\rm p}/\ell_{\rm ee}}$ for the $\nu_{\rm p}\neq 0$ case.

\begin{figure}[t!]
\centerline{\includegraphics[width=.99 \linewidth]{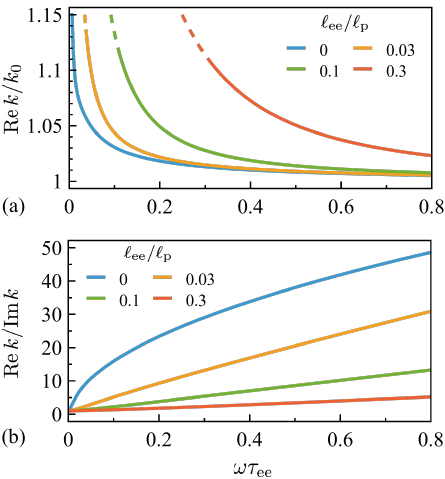}}
\caption{
Dispersion (a) and spatial quality factor (b) of propagating gapless acoustic plasmon mode in gated viscous 2DEF with weak momentum relaxation for a narrow channel geometry at {\rm Kn=0.05}. Dashed parts in (a) denote frequency range with $Q_{\rm k}(\omega)<1$ and $k_{0}$ is the wave vector of bulk acoustic plasmon.
\label{Fig4}
}
\end{figure}

Eq.~(\ref{Eq:Disp}) shows that properties of "viscous" kinetic inductor studied in Sec.~2 are translated into AP propagation, see Fig.~\ref{Fig4}. Namely, size effects in $\mathcal{L}_{\rm k}(\omega)$ (Fig.~\ref{Fig3}a) lead to non-linear and geometry dependent dispersion of AP, see Fig.~\ref{Fig4}a. The spatial quality factor of the plasmon wave $\mathcal{Q}(\omega)={\rm Re}\,k(\omega)/{\rm Im}\,k(\omega)$ due to particle collisions is determined solely by $Q_{\rm k}(\omega)$ (Fig.~\ref{Fig3}b). At small $Q_{\rm k}\ll 1$ it is close to unity $\mathcal{Q}\approx 1 + Q_{\rm k}$, while at $Q_{\rm k}\gg 1$ it is $\mathcal{Q}\approx 2 Q_{k}$. Fig.~\ref{Fig4}b shows that in 2DEF AP is stable to collisional damping and its propagation is described by relatively high quality factors once the condition of hydrodynamic transport is satisfied $\ell_{\rm ee}\ll \ell_{\rm p}, W$. In particular, at ${\rm Kn}=0.05$ and $\ell_{\rm ee}/\ell_{\rm p}=0.03$, $\mathcal{Q}=10$ is reached at $\omega\tau_{\rm ee}\approx 0.22$, which corresponds to 70~GHz in graphene ($\nu_{\rm ee}=2$ THz) and 7~GHz in GaAs ($\nu_{\rm ee}=0.2$ THz).

Renormalization of plasmon dispersion by enhanced $\mathcal{L}_{\rm k}(\omega)$ in narrow channel geometry (see Fig.~\ref{Fig4}a, solid curves) and $\mathcal{L}_{\rm k}(\omega)/\lambda_{\rm k}$ itself should be measurable via sensitive MW techniques~\cite{Yoon2014}. In Ref.~\cite{Yoon2014}, kinetic inductance measurements were performed at average $Q_{\rm k}=0.5$ with relative standard error for ${\rm Re}\,k$ about 7\%. Since in 2DEF $\mathcal{L}_{\rm k}$ enhancement domain overlaps with $Q_{\rm k}>1$ range, the rescaled measurement accuracy due to increasing $Q_{\rm k}$ shows that for ${\rm Kn}=0.05$ and $\ell_{\rm ee}/\ell_{\rm p}=0.03$, ${\rm Re}\,k/k_0$ stays close to threefold standard error within the relevant frequency domain of considerable $\mathcal{L}_{\rm k}(\omega)>\lambda_{\rm k}$. We note that this may obscure measurements of collective inertial mass in graphene~\cite{Yoon2014} extracted from $L_{\rm k}$ in hydrodynamic regime.


{\bf 4. Kinetic inductance in non-uniform parallel flow of 2D hydrodynamic electrons.} To address the origin of size effect in kinetic inductance and study the possibility to manipulate its values, we recall an energy argument, which is valid for arbitrary flow pattern and sample geometry. The general definition of $L_{\rm k}$ in 2DEF stems from representation of the total kinetic energy of an entire fluid as an energy stored in the equivalent inductor. For a quasi-one-dimensional AC parallel flow, e.g. the one considered in Sec.~2, this equivalence is given by $E_{\rm k}(t)=\frac{1}{2}L_{\rm k}I^2(t)$. To proceed further, one needs to specify the local kinetic energy density $\mathcal{E}_{\rm k}({\bf r},t)$ to obtain $E_{\rm k}(t)=\int_S \mathcal{E}_{\rm k}({\bf r},t)d{\bf r}$, where $S$ is sample area. In hydrodynamic regime, $\mathcal{E}_{\rm k}({\bf r},t)$ is directly linked to a local flow velocity ${\bf V}({\bf r},t)$ via quadratic form. Its coefficients are determined by dispersion law $\epsilon({\bf p})$ for non-interacting quasiparticles in a system. In the simplest case of isotropic and parabolic dispersion, kinetic energy density is $\mathcal{E}_{\rm k}=\rho_{\rm m}V^2/2$ with $\rho_{\rm m}=m N_0$ the mass density of fluid and $m$ the quasiparticle mass. This relation is conventional in classical fluid mechanics~\cite{Landau_V6}, however, among the variety of hydrodynamic materials, it is relevant only in GaAs nanostructures.

In this work, we consider the case of arbitrary isotropic quasiparticle dispersion and revisit expression for $\mathcal{E}_{\rm k}({\bf r},t)$ for it. Within the thermodynamics of a local equilibrium state, $\mathcal{E}_{\rm k}({\bf r},t)$ comes up for linear transport as a separate contribution to small perturbations of a thermodynamic potential density of fluid, which is associated with its motion. In the case of 2DEF, it is convenient to consider the grand potential density
\begin{multline}
    \label{Eq:Omega}
    \Omega({\bf r}, t)=g\sum\limits_{\bf p}\left[(\epsilon({\bf p})-\mu)n_{\bf p}+T n_{\bf p}\ln n_{\bf p}\right.+\\
    +\left.T(1-n_{\bf p})\ln (1-n_{\bf p})\right]\,,
\end{multline}
where $g$ and ${\bf p}$ are the degeneracy and momentum of quasiparticle state and $n_{\bf p}$ is given by
\begin{equation}
    \label{Eq:n_loc}
    n_{\bf p}({\bf r}, t)=\left[\exp\left(\frac{\epsilon({\bf p})-{\bf p}{\bf V}({\bf r},t)-\mu({\bf r},t)}{T({\bf r},t)}\right)+1\right]^{-1}\,.
\end{equation}
Distribution~(\ref{Eq:n_loc}) describes local equilibrium~\cite{Atwal2002, Narozhny2019} established by electron-electron scattering which conserves particle density, energy and momentum.

In a weakly perturbed state, deviation of~(\ref{Eq:n_loc}) from the absolute equilibrium distribution $n_0(\epsilon)$ with ${\bf V}({\bf r},t)=0$, $\mu({\bf r},t)=\mu_0$ and $T({\bf r},t)=T_0$ is small $\delta n_{\bf p}({\bf r},t)=n_{\bf p}({\bf r},t)-n_0(\epsilon)\ll 1$. Thus (\ref{Eq:Omega}) can be expanded by powers of $\delta n_{\bf p}$, $\delta T\ll T_0$ and $\delta \mu \ll \mu_0$. Taking into account that the Fermi-Dirac distribution minimizes~(\ref{Eq:Omega}), $\left.\delta \Omega/\delta n_{\bf p}\right|_{n_0}=0$, perturbation of the grand potential density to the second order in $\delta n_{\bf p}\ll1$ is given by $\delta\Omega[n_{\bf p}]\approx -s_0 \delta T-N_0\delta \mu - g\sum_{\bf p}\delta n_{\bf p}^2/2n'_0$. Here $s_0$ is the equilibrium entropy density. The first two terms in $\delta \Omega [n_{\bf p}]$ express the standard thermodynamic relation for a change of the grand potential at constant volume which holds locally. For the linearized distribution~(\ref{Eq:n_loc}), the remaining quadratic term yields the kinetic energy density 
\begin{equation}
    \label{Eq:Ek_density}
    \mathcal{E}_{\rm k}({\bf r},t)=\frac{1}{2}\rho_{\rm m}V^2({\bf r},t)=\frac{1}{2}\lambda_{\rm k}j^2({\bf r},t),    
\end{equation}
where $\rho_{\rm m}=-\frac{1}{2}\int_0^{+\infty}D(\epsilon) p^2(\epsilon) n'_0(\epsilon)d\epsilon$ is the inertial density of fluid, ${\bf j}({\bf r},t)=-eN_0 {\bf V}({\bf r},t)$ is the charge current density and $D(\epsilon)$ is the density of states. For $\epsilon(p)=p^2/2m$, the inertial density coincides with the mass density of a fluid $\rho_{\rm m}=m N_0$ proportional to the particle density $N_0$. In graphene, $\rho_{\rm m}=3\rho_{\epsilon}/2v_{\rm F}^2$ is proportional to the equilibrium energy density $\rho_{\epsilon}=\int_0^{+\infty}\epsilon D(\epsilon)n_0(\epsilon)d\epsilon$, instead of $N_0$. Using $\rho_{\rm m}$  one may introduce the collective mass $m=\rho_{\rm m}/N_0$ for the non-parabolic case as well~\cite{Yoon2014}.

Eq.~(\ref{Eq:Ek_density}) shows that for a non-uniform flow of 2DEF the local reactive behavior is still described by the Drude specific kinetic inductance $\lambda_{\rm k}=\rho_{\rm m}/e^2 N_0^2$. However, for the entire flow of 2DEF in a constrained geometry this local relation is violated as the total kinetic energy $E_{\rm k}(t)$ reduces to squared net current only if the flow is uniform. Eq.~(\ref{Eq:Ek_density}) and $E_{\rm k}(t)=L_{\rm k}I(t)^2/2$ yield, after averaging over the voltage period, the sheet kinetic inductance $\mathcal{L}_{\rm k}=W L_{\rm k}/l$ ($l$ is the sample length):
\begin{equation}
    \label{Eq:Lk}
    \mathcal{L}_{\rm k}=\lambda_{\rm k}\frac{\langle|V|^2\rangle}{|\langle V\rangle|^2}\,.
\end{equation}
Here $V(x)$ is the complex amplitude of the flow velocity and $\langle\ldots\rangle$ denotes averaging over the channel/constriction width. Eq.~(\ref{Eq:Lk}) shows that kinetic inductance of a collective flow in 2DEF is {\it always} greater than the one of individual non-interacting quasiparticles $\lambda_{\rm k}$. Relation $\mathcal{L}_{\rm k}\geq \lambda_{\rm k}$ implies that in hydrodynamic regime, collective inertia is increased since the adjacent layers of fluid are coupled by a transverse viscous flux of longitudinal momentum $\Pi_{xy}=-\rho_{\rm m}\eta\partial_x V$ and, therefore, experience mutual drag. In this sense, $\mathcal{L}_{\rm k}$ increase with respect to high-frequency ballistic value $\lambda_{\rm k}$ is similar to hydrodynamic superballistic conduction~\cite{Kumar2017} in stationary transport, with comparable enhancement magnitudes for the AC channel flow, see Fig.~\ref{Fig3}a.

The non-trivial behavior of $\mathcal{L}_{\rm k}(\omega)$ studied in Sec.~2 can be understood on the basis of Eq.~(\ref{Eq:Lk}). First, we note that Eq.~(\ref{Eq:Lk}) for the flow velocity profile~(\ref{Eq:V}) exactly reproduces the results presented in Fig.~\ref{Fig3}a. The decrease of $\mathcal{L}_{\rm k}$ with ${\rm Kn}$ and ${\rm Wo}$ in Fig.~\ref{Fig3}a is related to the flattening of $V(x)$ in each of these cases. Flow slip suppresses viscous dissipation and hence velocity profile non-uniformly throughout the channel by eliminating its trigger at the boundary. Viscous skin effect leads to non-uniform redistribution of viscous dissipation, predominantly near the boundaries. Therefore, $\mathcal{L}_{\rm k}$ enhancement decays rapidly with ${\rm Kn}$ and slowly with the Womersley number until the boundary viscous layers are developed at ${\rm Wo}\approx 15$, see Fig.~\ref{Fig2}. Effect fo ${\rm Gu}$ on $\mathcal{L}_{\rm k}(\omega)$ is qualitatively the same as ${\rm Wo}$.


{\bf 5. Discussion and conclusions.} It should be noted that expression similar to Eq.~(\ref{Eq:Lk}) describes size effect in kinetic inductance in superconductor structures~\cite{Meservey1969}. In this case the effect of $\mathcal{L}_{\rm k}$ enhancement is an attribute of large microscale systems and it is overpowered by increasing cross-section. In high-mobility 2DES, $\mathcal{L}_{\rm k}$ is enhanced simultaneously with decreasing cross-section and these effects are combined multiplicatively in the total sample inductance $L_{\rm k}=\mathcal{L}_{\rm k} l/W$.

High sensitivity of hydrodynamic transport to sample shape and boundary roughness can be exploited to manipulate viscous electronic flows via geometric engineering. This possibility has been widely studied in recent years~\cite{Keser2021, Levin2023, Varnavides2023, EstradaAlvarez2025_PRX, EstradaAlvarez2025_PRB} in the context of stationary flows and minimization of DC resistance. The present work highlights another potential application of hydrodynamic materials as mesoscale planar inductive elements in AC transport at MW frequencies. Their performance can be also tuned via geometric engineering, according to Eq.~(\ref{Eq:Lk}). This involves flow profile engineering aimed to maximize reactive response under the constrain of large kinetic inductor's quality factor $Q_{\rm k}$. For example, this can be done by introducing backflow patterns, especially into ${\rm Re}\,V$.

The authors are grateful to P.~S.~Alekseev for valuable discussions.


{\bf Funding.} This work was supported by the Russian Science Foundation project No. 25-22-00791.

{\bf Conflict of interest.} The authors of this work declare that they have no conflicts of interest.

\bibliography{LK_P_refs}

\end{document}